 \newcommand{\be}[0]{\begin{equation}}
 \newcommand{\ee}[0]{\end{equation}}
 \newcommand{\ba}[0]{\begin{eqnarray}}
 \newcommand{\ea}[0]{\end{eqnarray}}
\documentstyle[12pt, epsfig]{article}
\begin{document}
\Large
\hfill\vbox{\hbox{DTP-95/32}
            \hbox{June 1995}}
\nopagebreak

\vspace{2.75cm}
\begin{center}
\LARGE
{\bf Infrared Behaviour of the Gluon Propagator: Confining or Confined?}
\vspace{0.8cm}
\large

Kirsten B\"uttner

\vspace{0.5cm}
and
\vspace{0.5cm}

\large

M.R. Pennington

\vspace{0.4cm}
%\large
\begin{em}
Centre for Particle Theory, University of Durham\\
 Durham DH1 3LE, U.K.
\end{em}

\vspace{1.7cm}

\end{center}
\normalsize
\vspace{0.45cm}

%%%%%%%%%%%%ABSTRACT%%%%%%%%%%%%%%%%%%

{\leftskip = 2cm \rightskip = 2cm
The possible infrared behaviour of the gluon propagator is studied
analytically, using the Schwinger-Dyson equations, in both the axial
and the Landau gauge. The possibility of a gluon propagator less
singular than $1/k^{2}$ when $k^{2} \rightarrow 0$ is investigated and
found to be inconsistent, despite claims to the contrary, whereas an
infrared enhanced one is consistent. The implications for confinement
are discussed.
\par}

\newpage
\baselineskip=6.8mm
\parskip=2mm
{\bf {\large 1. Introduction}}
\vspace{0.3cm}

The gluon propagator $\Delta_{\mu\nu}(k)$ is gauge dependent and as
such is not experimentally observable. However it's infrared behaviour
has important implications for quark confinement.
It can be shown, that a gluon propagator, which is as singular as
$1/k^{4}$ when $k^{2} \rightarrow 0$ indicates that the interquark
potential rises linearly with the separation. More formally, West
\cite{West} proved, that if, in any gauge,
$\Delta_{\mu\nu}$ is as singular as $1/k^{4}$ then the Wilson operator
satisfies an area law, often regarded as a signal for confinement.
Hence, quarks are confined through gluon interaction.

Another sufficient condition for confinement is, that a propagator
of a coloured state should not have any singularities on the real,
positive $k^{2}$-axis \cite{RWK}. So, if gluons are confined, then they
cannot propagate on-shell and $\Delta_{\mu\nu}$ must be less singular
than $1/k^{2}$ when $k^{2} \rightarrow 0$. Such a behaviour of the
gluon propagator was first assumed by
Landshoff and Nachtmann \cite{LN} on purely phenomenological grounds,
being needed to reproduce experiment with their model of the pomeron.

Clearly (Fig.~1) the gluon propagator cannot both be more singular
and less singular than $1/k^2$ as $k^2\to 0$, but which is correct~?
The Schwinger-Dyson equations provide the natural starting point
for a non-perturbative investigation of this infrared behaviour of the
gluon propagator. Extensive work has been previously performed in both
the axial gauge \cite{BBZ}--\cite{CR} and the Landau gauge
\cite{Mandel}--\cite{BP}.
(For a comprehensive review see Roberts and Williams~\cite{SDE}.)
A solution as singular as $1/k^{4}$ has been shown to exist in both
gauges \cite{BBZ}--\cite{Schoen} and \cite{Mandel}--\cite{BP},
whereas a {\it confined} solution for the gluon propagator, i.e. less singular
than $1/k^{2}$, has only been claimed to exist in the axial gauge \cite {CR}.
The purpose of this paper is to explore why these two different
behaviours have been found. Fortunately, in studying just the infrared
behaviour, there is no need to solve the Schwinger-Dyson equation at
all momenta. It is this that greatly simplifies our discussion and
allows an analytic treatment.

In Sect.~2 we briefly describe the Schwinger-Dyson equation for the
gluon propagator. The axial gauge studies are reviewed in Sect.~3 and
the possible, self-consistent solutions for the infrared behaviour of
the gluon propagator are reproduced analytically. In Sect.~4 we repeat
the discussion for the Landau gauge and find, that a propagator less
singular than $1/k^{2}$ when $k^{2} \rightarrow 0$ is not a solution of
the Schwinger-Dyson equation. In Sect.~5 we discuss the differing forms
of the Schwinger-Dyson equations used to deduce these results.
In Sect.~6 we state our conclusions. \\
\newpage
\baselineskip=7mm

\begin{figure}[h]
\begin{center}
{}~\epsfig{file=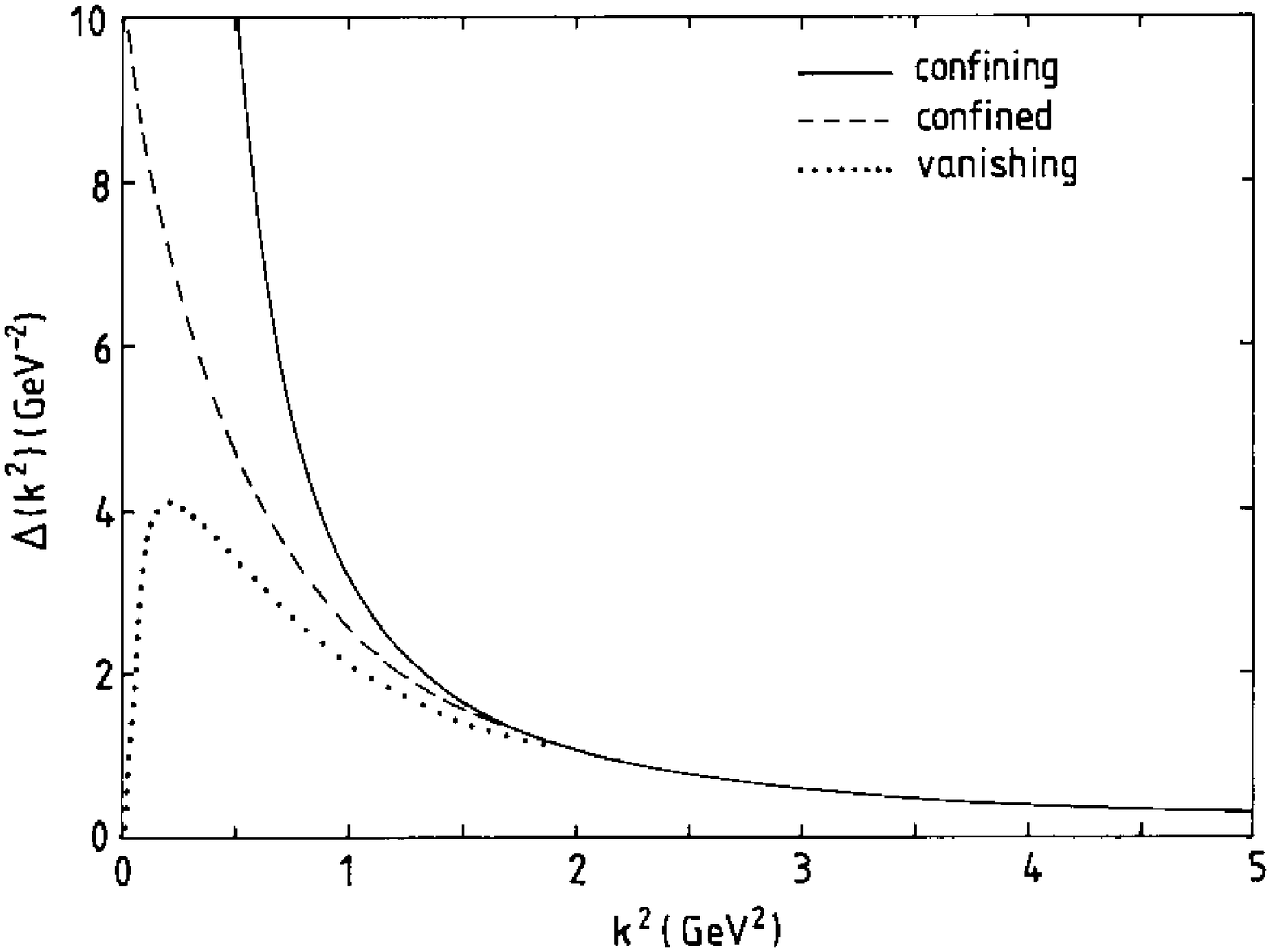,angle=90,width=13cm}
\end{center}
\caption[qcd2]{Possible behaviour of the gluon propagator $\Delta(k^2)$, which
is the coefficient of the $g_{\mu \nu}$ or $\delta_{\mu \nu}$
component of $\Delta_{\mu\nu}(k)$.\\
%\begin{itemize}
 (a) {\it confining} gluon, $\Delta \sim (k^2)^{-2}$,\\
 (b) {\it confined} gluon, $\Delta \sim (k^2)^{-c}$ with $c$ very small,\\
 (c) infrared vanishing gluon $\Delta \sim k^{2}$.\\
%\end{itemize}
All are matched to the perturbative behaviour for $k$ larger than a few GeV.}
\end{figure}
\newpage

\vspace{1.5cm}
{\bf {\large 2. Schwinger-Dyson equation for the gluon propagator}}
\vspace{0.5cm}

The Schwinger-Dyson equations are coupled integral equations which
inter-relate the Green's functions of a field theory. Since they build
an infinite tower of coupled equations, approximations and truncations
are necessary to solve them.
The Schwinger-Dyson equation for the gluon propagator yields a relation
for $\Delta_{\mu\nu}$ in terms of the full 3 and 4-point vertex
functions, $\Gamma_{\mu\nu\rho}^{3}$ and $\Gamma_{\mu\nu\rho\sigma}^{4}$,
the quark and the ghost propagators and couplings.
The equation is displayed diagrammatically in Fig.~2.
Here we only consider a pure gauge theory, i.e. a world without quarks.
This is reasonable since  we expect it is the non-Abelian nature of QCD
which is responsible for confinement.

\begin{figure}[h]
\begin{center}
{}~\epsfig{file=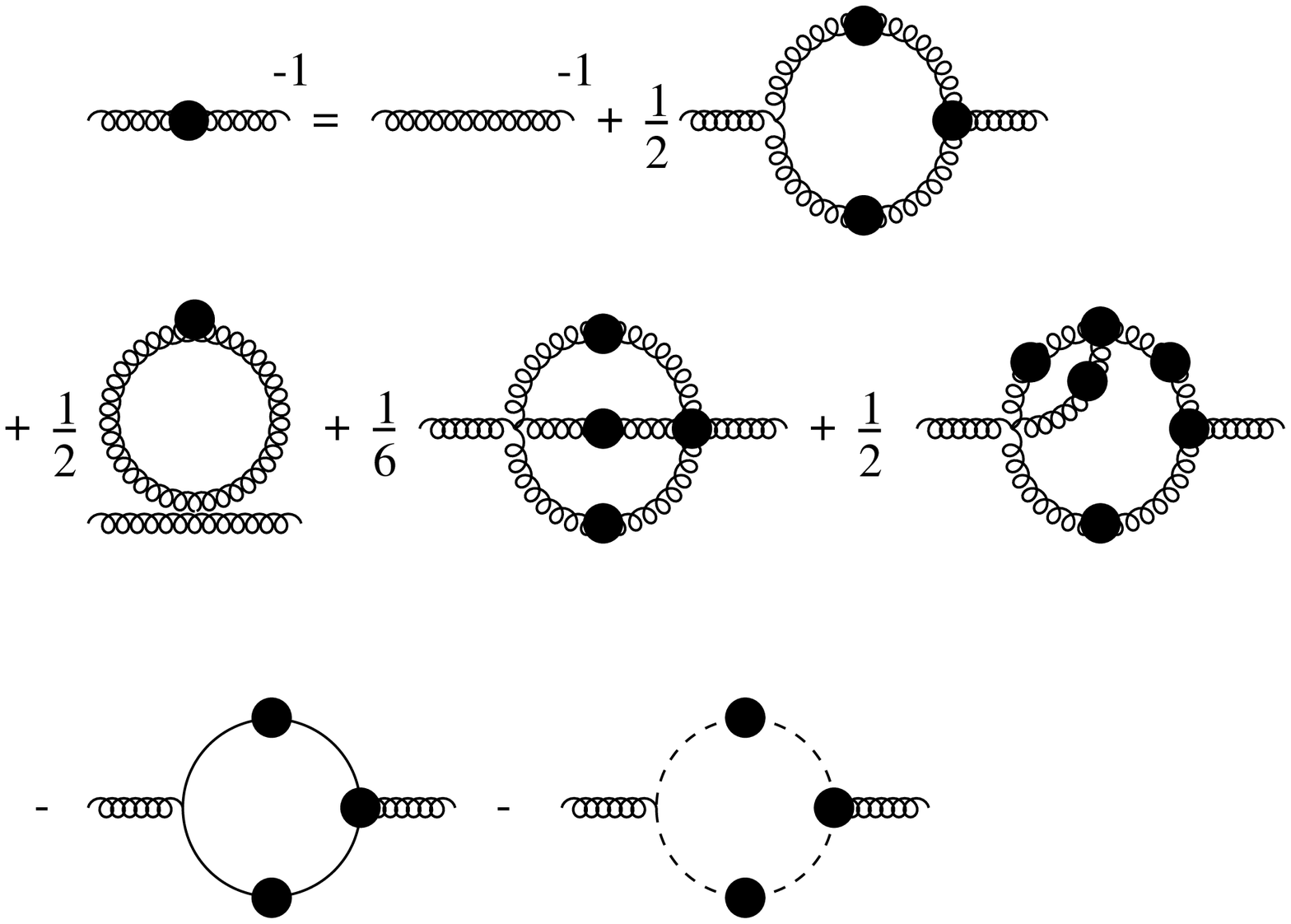,width=13cm}
\end{center}
\caption[qcd1]{The Schwinger-Dyson equation for the gluon propagator\\
Here the broken line represents the ghost propagator. The symmetry
factors 1/2 and 1/6 and a negative sign for every ghost and fermion loop
arise from the usual Feynman rules.}
\end{figure}

\newpage
\baselineskip=7.2mm

\vspace{0.8cm}
{\bf {\large 3. Axial Gauge Studies}}
\vspace{0.4cm}

In the axial gauge the gluon propagator is transverse  to the gauge
vector $n_{\mu}$, so
\begin{center}
Axial gauge formalism : $n_{\mu}A^{a\mu} = 0$
\end{center}
\noindent Studies of the axial gauge Schwinger-Dyson equation have the
advantage that
ghost fields are absent and the four-gluon vertex terms, Fig.~2, may be
projected out
of the Schwinger-Dyson equation. However they have the drawback that the
gluon
propagator depends not only on $p^{2}$, but also on the unphysical gauge
parameter $\gamma$, defined as
\[ \gamma = \frac{(n\cdot p)^{2}}{n^{2}p^{2}}\]
and in general must depend on two scalar functions, $F$ and $H$~:
\be
 \Delta_{\mu\nu}(p^{2},\gamma) = -\frac{i}{p^{2}}\left[F(p^{2},\gamma)
  M_{\mu\nu} + H(p^{2},\gamma) N_{\mu\nu}\right] \quad ,
\ee
 with the tensors given by:

\[M_{\mu\nu} = g_{\mu\nu} - \frac{p_{\mu}n_{\nu}+p_{\nu}n_{\mu}}
{n\cdot p} + n^{2}\frac{p_{\mu}p_{\nu}}{(n\cdot p)^{2}} \qquad ,\]

\[N_{\mu\nu} = g_{\mu\nu} - \frac{n_{\mu}n_{\nu}}{n^{2}} \qquad .\]

\noindent The free propagator is obtained by substituting $F=1$ and $H=0$.
\vspace{0.3cm}

In all previous axial gauge studies it has been assumed that any infrared
singular part of
the propagator has the same tensor structure as the free one (though
importantly this contradicts the results of West \cite{West2}).
Thus for $p^{2} \rightarrow 0$ it is assumed that
\be
\Delta_{\mu\nu}(p^{2},\gamma) = -\frac{i}{p^{2}}\,F(p^{2},\gamma) M_{\mu\nu}
 \qquad .
\ee

\noindent The gluon vacuum polarization tensor $\Pi_{\lambda\mu}
(p^2,\gamma)$ is defined by
\[\Pi^{\lambda\mu} \ \Delta_{\mu\nu} = g^{\lambda}_{\ \nu} -
\frac{n^{\lambda} p_{\nu}}{n \cdot p} \qquad. \]

\newpage

\noindent Projecting the integral equation with ${n_{\mu}n_{\nu}}
/{n^{2}}$ the loops involving the four gluon vertex give an identically
zero contribution because of  the tensor structure
of the bare 4-gluon vertex and the fact that the gluon propagator is
transverse to the axial gauge vector, that is
\[n_{\mu}\Delta^{\mu\nu} = 0 = \Delta^{\mu\nu}n_{\nu} \qquad . \]

\noindent Thus the relevant part of the Schwinger-Dyson equation of
Fig.~2 becomes:
\ba
\Pi_{\mu\nu} = \Pi_{\mu\nu}^{(0)} &-&\frac{g^2}{2}\int\frac{d^{4}k}
{(2\pi)^{4}}\ \Gamma_{\mu\alpha\delta}^{(0)}(-p,k,q)\
\Delta^{\alpha\beta}(k)\ \Delta^{\gamma\delta}(q)\ \Gamma_{\beta\gamma
\nu}(-k,p,-q)\ \nonumber\\
&-& \frac{g^2}{2}\int\frac{d^{4}k}{(2\pi)^{4}}\ \Gamma_{\mu\nu\alpha
\beta}^{0}(p,k,-k,-p)\ \Delta^{\alpha\beta}(k)
\ea
where $q = p-k$, the last term is the tadpole contribution and all colour
indices are implicitly included in the vertices.
Once the full 3-gluon vertex is known, we have a closed equation for the
gluon vacuum polarization $\Pi_{\mu\nu}$.

The vertex is constrained by the Slavnov-Taylor identity in terms of this
vacuum polarization~:
\be
q_{\nu}\Gamma^{\mu\nu\rho}(p,q,k) = \Pi^{\mu\rho}(k) - \Pi^{\mu\rho}(p)
\qquad .
\ee
Separating $\Gamma^{\mu\nu\rho}$ into transverse and longitudinal part,
where the transverse part is defined to vanish when contracted with any
external momentum, the Slavnov-Taylor identity exactly determines the
longitudinal part \cite{BC} if it is to be free of kinematic
singularities. Thus
\ba
\Gamma_{\mu\nu\rho}^{L}(p,k,q) &=& g_{\mu\nu} \left(\frac{p_{\rho}}
{F(p^{2},\gamma)} - \frac{q_{\rho}}{F(q^{2},\gamma)}\right) \nonumber \\
& & + \frac{1}{p^{2}-q^{2}}\left(\frac{1}{F(p^{2},\gamma)} -
\frac{1}{F(q^{2},\gamma)}\right)(p_{\nu}q_{\mu} - g_{\mu\nu}p\cdot q)
(p_{\rho}-q_{\rho})
\nonumber
 \\ & & +\phantom{\frac{1}{p^2}}\ \mbox{cyclic permutations}\qquad .
\ea

This longitudinal part is responsible for the dominant ultraviolet structure of
the vertex. Moreover, it is assumed, that it entirely embodies the
infrared behaviour, and so the transverse part can be neglected. This
assumption is motivated by the fact that the transverse part (as defined)
vanishes, when the external momenta approach zero.

Using the explicit expressions, Eq.~(2) for $\Delta$, Eqs.~(4,5) for
$\Gamma$ and multiplying with ${n_{\mu}n_{\nu}}/{n^{2}}$, we find in
Euclidean space:

\ba
-\frac{p^{2}}{F(p^{2})}\left(1-\gamma\right) &=& -p^{2}\left(
1-\gamma\right) +
\frac{g^{2}C_{A}}{2}\int\frac{d^{4}k}{(2\pi)^{4}}\frac{n\cdot
(k-q)}{n^{2}}
\Delta^{\lambda\rho}_{(0)}(k)\ \Delta^{\lambda\sigma}_{(0)}(q)
\nonumber \\
& &\left\{k\cdot n\left[\delta_{\rho\sigma}F(q^{2})
 -\frac{F(q^{2})-F(k^{2})}
{k^{2}-q^{2}}(\delta_{\rho\sigma} k\cdot q -k_{\rho}q_{\sigma})
\right. \right.
\nonumber \\
& & \qquad \qquad \qquad
\left. +\frac{ F(k^{2})-F(p^{2})}{p^{2}-k^{2}}\frac{F(q^{2})}
{F(p^{2})} p_{\sigma}(p+k)_{\rho}\right]\nonumber \\
& & - q\cdot n\ \left[\delta_{\rho\sigma}F(k^{2})-\frac{F(q^{2})-
F(k^{2})}{k^{2}-q^{2}}(\delta_{\rho\sigma} k\cdot q -
k_{\rho}q_{\sigma}) \nonumber \right.\\
& & \qquad \qquad \qquad
\left.\left. +\frac{ F(q^{2})-F(p^{2})}{p^{2}-q^{2}}\frac{F(k^{2})}
{F(p^{2})}p_{\rho}(q+p)_{\sigma}\right] \right\} \nonumber \\
& & + \frac{g^{2}C_{A}}{2}\int\frac{d^{4}k}{(2\pi)^{4}} \frac{F(k^{2})}
{k^2}\left( 2 + \frac{k^2n^2}{(n\cdot k)^2} \right) \qquad .
\ea

\noindent This is the equation first found by Baker, Ball and
Zachariasen \cite{BBZ} who studied its solution numerically.
They came to the conclusion that the only consistent infrared behaviour
for the function $F(p^{2})$ is
\[ F(p^{2}) \propto \frac{1}{p^{2}} \ \ \ \mbox{as}\ p^{2} \
\rightarrow 0\]
and that this is independent of $\gamma$ as a numerical approximation.

Schoenmaker \cite{Schoen} simplified the BBZ equation further by exactly
setting $\gamma = 0$. Doing this the contribution of the tadpole
diagram vanishes. Moreover, approximating $F(q^2)$ by $F(p^2+k^2)$, which
should be exact in the infrared limit, allows the angular integrals to
be performed analytically.
Consequently, Schoenmaker finds the following simpler equation:
\ba
\lefteqn{p^2\left(\frac{1}{F(p^2)} - 1\right) = } \nonumber \\
& &\frac{g^{2}C_{A}}{32\pi^2}\left\{\int_{0}^{p^2} dk^2 \left[\left(
-\frac{k^4}{12p^4}+\frac{5}{2}\frac{k^2}{p^2}-\frac{2}{3}\frac{k^2}{p^2-k^2}
\right)F_{1} +\left(\frac{k^6}{24p^6}-\frac{1}{4}\frac{k^4}{p^4}
 -\frac{1}{4}\frac{k^2}{p^2}\right)F_{2} \nonumber \right.\right.\\
& & \qquad \quad \left.
+ \left(-\frac{1}{6}\frac{k^2}{p^2}+\frac{2}{3}\frac{k^2}{p^2-k^2}
\right)F_{3}\right]
+ \int_{p^2}^{\infty} dk^2 \left[\left( -\frac{3}{4}-\frac{p^2}{4k^2}-
\frac{2}{3}\frac{p^2}{p^2-k^2}\right)F_{1} \nonumber \right. \\
& & \qquad \quad \left.\left. +\left(-\frac{3}{4}\frac{k^2}{p^2}+\frac{5}{12}
-\frac{1}{8}\frac{p^2}{k^2}\right)F_{2} + \left(\frac{7}{6}
\frac{p^2}{k^2}+\frac{2}{3}\frac{p^2}{p^2-k^2}\right)F_{3}\right]
\right\} \qquad ,
\ea

where
\ba
 F_{1} &=& F(p^2+k^2) \nonumber \qquad\qquad ,\\
 F_{2} &=& F(p^2+k^2)-F(k^2) \nonumber \quad ,\\
 F_{3} &=& \frac{F(p^2+k^2)F(k^2)}{F(p^2)} \qquad .\nonumber
\ea

\noindent In general, this equation has a quadratic ultraviolet divergence,
which
would give a mass to the gluon. Such terms have to be subtracted to ensure
the masslessness condition
\be \lim_{p^2 \rightarrow 0} \Pi_{\mu\nu} = 0 \ \ \mbox{, i.e.}\ \ \
\frac{p^2}{F(p^2)} = 0 \ \ \mbox{for}\ \ p^2 \rightarrow 0 \qquad ,\ee
is satisfied. This property can be derived generally from the Slavnov-Taylor
identity and always has to hold.

The complicated structure of the integral equation, Eq.~(7),
does not allow an exact analytic solution for the gluon renormalization
function $F(p^2)$ to be found and most previous studies (\cite{Mandel},
\cite{BP}, \cite{BBZ} and \cite{CR}) solve the equation numerically.
However, the possible asymptotic behaviour of $F(p^2)$ for both small
and large $p^2$ can be investigated analytically.

We determine which infrared behaviour of $F(p^2)$ can give a
self-consistent solution to the integral equation by taking a trial
input function $F_{in}(p^2)$ and substituting it into the right hand
side of the equation. After performing the $k^2$-integration, we obtain
an output function $1/F_{out}(p^2)$ to be compared to the reciprocal
of the input function.
To do this, the gluon renormalization function is approximated in the
infrared region by a Laurent expansion in powers of $p^2$ and at large
momenta by its bare form, i.e.
\be F(p^2) = \left\{ \begin{array}{ll}
\sum_{n=0}^{\infty} a_{n} \left(p^2/\mu^2 \right)^{n+\eta}
& \mbox{for \ $p^2 < \mu^2$}  \\ \\
1 & \mbox{for \ $p^2 > \mu^2$}
\end{array}
\right. \qquad ,\ee

\noindent where
\[\sum_{n=0}^{\infty} a_{n} = 1\] to ensure continuity at $p^2 =
\mu^2$. $\mu$ is the mass scale above which we assume perturbation
theory applies. $\eta$ can be negative to allow for an infrared enhancement.
Eq.~(9) is a sufficiently general representation for
finding the dominant self-consistent infrared behaviour. Of course, the
true renormalization function is modulated by powers of logarithms of
momentum, characteristic of a gauge theory. However, these do not
qualitatively affect the dominant infrared behaviour and can be neglected.
Indeed to make the presentation straightforward, we only need approximate
$F(p^2)$ by its dominant infrared power $(p^2)^{\eta}$ for $p^2 < \mu^2$
to test whether consistency is possible and this is what we describe
below. However, as we shall see, if $\eta$ is negative then potential
mass terms arise and these have to be subtracted. Only in this case do
higher terms in Eq.~(9) play a role too and it is necessary to consider
other than the leading term in the low momentum input. Otherwise higher
powers make no qualitative difference as we have checked. Consequently
we present only the results with the lowest powers in the representation,
Eq.~(9).

To illustrate the idea, let us take the trial infrared behaviour to be
just

\be F(p^2)\propto \left(\frac{p^2}{\mu^2}\right)^{\eta} \qquad \mbox{(i.e.
$a_{n} = 0$ for $n \geq 1$)} \qquad .\ee
Note that the masslessness condition, Eq.~(8), restricts $\eta$ to be less
than 1.
Furthermore we demand that in the high momentum region the solution of
the integral equation matches the perturbative result, i.e. for $p^2
\rightarrow \infty$, we have $F(p^2) = 1$, modulo logarithms.

\vspace{0.4cm}

Taking $\eta = -1$, for example, i.e.
\[F_{in}(p^2) = A\frac{\mu^2}{p^2}\]
in Schoenmaker's approximation, Eq.~(7), gives

\[p^2\left(\frac{1}{F(p^2)} - 1\right) = \mbox{const} \qquad . \]

\noindent This violates the masslessness condition of Eq.~(8) and so
has to be mass renormalized. As explained above, now terms in $F(p^2)$
of higher order in $p^2$ will generate a contribution to the right hand
side of the equation making it possible to find a self-consistent
solution by these cancelling the explicit factor of 1.
Consequently, we can approximate Eq.~(9) by
\be
F_{in}(p^2) = \left\{ \begin{array}{ll}
A \left({\mu^2}/{p^2} \right) + \left({p^2}/{\mu^2} \right)
& \mbox{if $p^2 < \mu^2 $} \\ \\
1 & \mbox{if $p^2 > \mu^2 $}
\end{array}
\right. \qquad ,
\ee

We then find, after mass renormalization:
\be
\frac{1}{F_{out}(p^2)} = 1 + \frac{g^{2}C_{A}}{32\pi^2} \left[
\frac{67}{96}\frac{p^2}{\mu^2} - \frac{1}{12}
 -\frac{3}{8}\frac{p^2}{\mu^2}\ln\left(\frac{\mu^2}{p^2}\right) +
\frac{5}{12} \ln \left(\frac{\Lambda^2}{\mu^2}\right) \right] \qquad ,
\ee
\noindent where $\Lambda$ is the ultraviolet cut-off introduced to make
the integrals finite.
The ultraviolet divergent constant can be arranged to cancel the 1 and
we find self-consistency modulo logarithms. It is this result that
Schoenmaker found \cite{Schoen} supporting the earlier result of BBZ
\cite{BBZ}. However, importantly, self-consistency requires $A$, Eq.~(11),
to be
negative as also found by Schoenmaker.

\vspace{0.4cm}

More recently, Cudell and Ross \cite{CR} have taken Schoenmaker's
equation, Eq.~(7), and investigated whether one can find
self-consistency for a gluon renormalization function which is
less singular than $1/k^2$ for $k^2 \rightarrow 0$, i.e. which
corresponds to {\it confined} gluons.
The trial input function they use in their investigation is
\[F_{in}(p^2) \propto (p^2)^{1-c}\qquad ,\]
where c is small and positive to ensure a massless gluon, Eq.~(8).
Once more we want the integral equation for $\Pi_{\mu\nu}$ to agree with
perturbation theory in the ultraviolet region, but $F_{in}(p^2) \propto
(p^2)^{1-c}$ grows for large momenta and hence spoils the ultraviolet
behaviour. So to check whether this input function gives
self-consistency in the infrared, we input the  trial form~:
\be
F_{in}(p^2) = \left\{ \begin{array}{ll}
\left({p^2}/{\mu^2}\right)^{1-c} & \mbox{if $p^2<\mu^2$} \\ \\
1 & \mbox{if $p^2>\mu^2$}
\end{array}
\right. \qquad .
\ee
\noindent Inserting this into Eq.~(7), we find, after mass
renormalization~:
\ba
\frac{1}{F_{out}(p^2)} &=& 1 + \frac{g^2C_{A}}{32\pi^2} \left[
\ D_{1} + \ D_{2} \left( \frac{\mu^2}{p^2}\right)^{1-c}
+ \ D_{3} \left(\frac{p^2}{\mu^2} \right)^{1-c} +\  D_{4}\left(
\frac{p^2}{\mu^2}\right)^{c} + ...\right],
\ea
\noindent where $F_{1}, F_{2}$ and $F_{3}$ have been expanded
for small $p^2$ and only the first few terms have been collected in this
equation so that
\ba
D_{1} &=& \frac{5}{12(1-c)} - \frac{1}{3} + \frac{11}{12}
\ln{\frac{\Lambda^2}{\mu^2}}  \nonumber \qquad ,\\
D_{2} &=& \frac{1}{2(2-2c)} + \frac{1}{6} \ln \left( \frac{\Lambda^2}
{\mu^2} \right) \nonumber \qquad ,\\
D_{3} &=& \frac{2407}{1440} - \frac{353c}{360} - \frac{3}{8c}
-\frac{1}{24(5-c)} + \frac{1+2c}{12(4-c)} - \frac{7-8c}{12(3-c)}
-\frac{7+c}{12(2-c)} -\frac{7}{12(1-c)} \nonumber \\
& & + \frac{1-4c}{6(2-2c)} + \frac{3-7c}{6(1-2c)} +
\frac{2}{3}(2-c)\Psi(1) - \frac{4}{3}(2-c)\Psi(1-c) + \frac{2}{3}
(2-c)\Psi(1-2c) \nonumber \ ,\\
D_{4} &=& - \frac{7-9c}{24c} \nonumber \qquad ,
\ea

\noindent where $\Psi$ is the logarithmic derivative of the Gamma
function.
Again the 1 can be arranged to cancel with the constant term and the
dominant infrared behaviour is indeed

\be \frac{1}{F_{out}(p^2)} \rightarrow \left(\frac{\mu^2}{p^2}\right)
^{1-c} \ \ \ \ \mbox{for} \ \ p^2 \rightarrow 0\ee

\noindent Hence a gluon propagator less singular than $1/p^2$ for $p^2
\rightarrow 0$ can be derived from Schoenmaker's equation as Cudell and
Ross \cite{CR} have found. Note once again that terms of higher order in
Eq.~(9)
do not qualitatively alter the result. Thus we see in the axial gauges
that apparently both {\it confined} and {\it confining} solutions are
possible for the gluon propagator. However, the singular {\it confining}
behaviour must be an artefact of the approximation that one of the gluon
functions, $H$, vanishes, since West \cite{West2} has shown that in a gauge
with only positive norm-states, a singular gluon renormalization function
 is not possible.
Moreover, the approximation of setting $\gamma = 0$ in the BBZ-equation,
Eqs.~(6,7),
has been seriously questioned in Ref.~\cite{Atkinson}.
It is therefore sensible to ask what is the behaviour in covariant gauges,
to which we now turn.

\vspace{0.8cm}
{\bf {\large 4. Landau Gauge Studies}}
\vspace{0.4cm}

The advantage of Landau gauge studies is the much simpler structure of
the gluon propagator, which is defined by~:
\be
\Delta_{\mu\nu} = - i\ \frac{G(p^2)}{p^2}\left(g_{\mu\nu} - \frac{p_{\mu}
p_{\nu}}{p^2}\right)\qquad .
\ee
\noindent However other problems arise and the following approximations
have to be made~:\\
\begin{itemize}
\item In any covariant gauge, ghosts are necessary to keep the vacuum
polarization transverse and hence are present in the Schwinger-Dyson
equation of the gluon propagator, Fig.~2. However, in all previous
studies~\cite{Mandel,BP} the ghost loop diagram is only included in as
much as it ensures the transversality of the gluon propagator, assuming
that otherwise it does not affect the infrared behaviour of the
propagator. This assumption is supported by the fact, that in a one-loop
perturbative calculation the ghost loop makes a numerically small
contribution to $G(p^2)$.
\item  The 4-gluon terms cannot be eliminated as in the axial gauge and
are simply neglected. This can be regarded as a first step in a
truncation of the Schwinger-Dyson equations.\\
\end{itemize}

With these assumptions, we again find a closed integral equation for the
gluon vacuum polarization, $\Pi_{\mu\nu}$, once the full 3-gluon vertex
is known.
In the Landau gauge, the Slavnov-Taylor identity for the 3-gluon vertex
involves the ghost self-energy, which is simply set to zero, and the
proper ghost-gluon vertex function $G_{\mu\nu}$.
However, in the limit of vanishing ghost momentum the ghost-gluon vertex is
approximately equal to the gluon propagator.
With this simplification the Slavnov-Taylor identity has the same form
as in the axial gauge and is given in Eq.~(5).
Once again neglecting the transverse part of the vertex, we obtain a
closed integral equation~:
\be
\Pi_{\mu\nu} = \Pi_{\mu\nu}^{(0)} - \frac{g^2}{2}\ \int\frac{d^{4}k}
{(2\pi)^{4}}\ \Gamma_{\mu\alpha\delta}^{(0)}(-p,k,q)\ \Delta^
{\alpha\beta}(k)\ \Delta^{\gamma \delta}(q)\ \Gamma_{\beta\gamma\nu}
(-k,p,-q)
\ee
\noindent where once again the colour indices are implicit and $q = p-k$.

A scalar equation  is obtained by projecting with
\[P^{\mu\nu} = \frac{1}{3p^2}(4p^{\mu}p^{\nu} - p^2g^{\mu\nu})\]

\noindent This projector has the advantage that the $g_{\mu \nu}$ term in
Eq.~(16) that is quadratically divergent in 4-dimensions, does not
contribute. Thus we find
\ba
\lefteqn{\frac{1}{G(p^2)} =
1 + \frac{g^{2}C_{A}}{96\pi^4}\frac{1}{p^2}\int d^4k \left[ G(q^2)
A(k^2,p^2) + \frac{G(k^2)G(q^2)}{G(p^2)}\ B(k^2,p^2)\right.} \nonumber \\
& & \qquad \quad + \left.\frac{G(k^2)-G(p^2)}
{k^2-p^2}\frac{G(q^2)}{G(p^2)}\ C(k^2,p^2) + \frac{G(q^2)-G(k^2)}
{q^2-k^2} \ D(k^2,p^2) \right] ,
\ea
\noindent where
\ba
 A(k^2,p^2) &=& 48\frac{(k\cdot p)^2}{k^2p^2q^2} - 64\frac{(k\cdot p)}
{k^2q^2} + 16\frac{(k\cdot p)^3}{k^2p^2q^4} - 12\frac{1}{k^2} + 22
\frac{p^2}{k^2q^2} - 42\frac{(k\cdot p)^2}{k^2q^4}
- 10\frac{p^4}{k^2q^4} \nonumber \\
& & + 36\frac{p^2(k\cdot p)}{k^2q^4}\qquad , \nonumber \\
B(k^2,p^2) &=& - 13\frac{p^2}{q^4} + 18\frac{p^2(k\cdot p)}{k^2q^4}
- 2\frac{(k\cdot p)^2}{k^2q^4} - 4\frac{p^4}{k^2q^4} + \frac{p^2
(k\cdot p)^2}{k^4q^4}\; , \nonumber \\
C(k^2,p^2) &=& 4\frac{(k\cdot p)^2}{k^2q^2} + 6\frac{p^2(k\cdot p)}
{k^2q^2} + 6\frac{(k\cdot p)}{q^2} + 8\frac{p^2}{q^2}
\quad ,  \nonumber \\
D(k^2,p^2) &=& 12\frac{k^2}{q^2} - 48\frac{(k\cdot p)^2}{p^2q^2} + 48
\frac{(k\cdot p)^3}{k^2p^2q^2} +24\frac{(k\cdot p)}{q^2} -5\frac{p^2}
{q^2} - 40\frac{(k\cdot p)^2}{k^2q^2} +9\frac{p^2(k\cdot p)}{k^2q^2}
\, .\nonumber
\ea

Brown and Pennington \cite{BP} studied this equation numerically and
found

\[G(p^2) =  A\, \frac{\mu^2}{p^2} \ \ \ \ \mbox{for} \ \ p^2 \rightarrow 0\]

\noindent to be a consistent solution. This result is in agreement with
Mandelstam's study of the gluon propagator~\cite{Mandel}.

\vspace{0.4cm}

Again approximating $G(q^2)$ by $G(p^2+k^2)$ allows us to perform the
angular integrals analytically in Eq.~(18), giving:
\ba
\lefteqn{\frac{1}{G(p^2)} = 1 + } \nonumber \\
& &\frac{g^{2} C_{A}}{48\pi^2}\frac{1}{p^2}\left\{
\int_{0}^{p^2} dk^2 \left[ G_{1} \left( -1 - 10\frac{k^2}{p^2}
+6\frac{k^4}{p^4} + \frac{k^2}{p^2-k^2} \left( \frac{75}{4} -
\frac{39}{4}\frac{k^2}{p^2} + 4\frac{k^4}{p^4} - 5\frac{p^2}{k^2}\right)
\right) \right.\right. \nonumber \\
& & \left. \qquad \qquad + G_2 \left(-\frac{21}{4}\frac{k^2}{p^2} + 7
\frac{k^4}{p^4} - 3 \frac{k^6}{p^6}\right) +
G_3 \left( \frac{k^2}{p^2-k^2}\left( -\frac{27}{8} - \frac{11}{4}
\frac{k^2}{p^2} - \frac{15}{8}\frac{p^2}{k^2} \right)\right)\right]
\nonumber \\
& & \qquad \qquad +\int_{p^2}^{\infty} dk^2 \left[ G_{1} \left( \frac{p^2}{k^2}
- 6 +
\frac{p^2}{p^2-k^2}\left( \frac{29}{4} + \frac{3}{4}\frac{p^2}{k^2}
\right)\right) \,+ \right.\nonumber \\
& & \qquad \qquad \left.\left. + G_2 \left( - \frac{3}{2} + \frac{1}{4}
\frac{p^2}{k^2} \right) +
G_3 \left( \frac{p^2}{p^2-k^2}\left( \frac{3}{4} -
\frac{67}{8}\frac{p^2}{k^2} - \frac{3}{8}\frac{p^4}{k^4} \right)\right)
\right] \right\}\, ,
\ea
where
\ba
G_{1} &=& G(p^2+k^2) \qquad\qquad ,\nonumber \\
G_{2} &=&  G(p^2+k^2)- G(k^2) \quad ,\nonumber \\
G_{3} &=& \frac {G(k^2)G(p^2+k^2)}{G(p^2)} \; . \nonumber
\ea

Note that the integral equation has the usual ultraviolet divergences,
but infrared divergences are also possible. The ultraviolet divergences
can be handled in the standard way to give a renormalized function
$G_{R}(p^2)$ --- this will not be discussed here. However we have to
make the potentially infrared divergent integrals finite in order to
calculate the integrals\footnotemark.
\footnotetext{These divergences do not arise in an axial gauge when
$\gamma$ is set equal to zero as Schoenmaker does, Eq.~(7).}
The infrared regularization procedure proposed by Brown and Pennington
\cite{BP}
is to use the plus prescription of the theory of distributions,
which is defined as follows:

\[ \left( \frac{\mu^2}{k^2} \right)_{+} = \frac{\mu^2}{k^2} \ \ \ \ \
\mbox{for} \ \ \infty > k^2 > 0 \]

\noindent and in the neighbourhood of $k^2 = 0$ it is a distribution
that satisfies:
\ba
\lefteqn{\int_{0}^{\infty} dk^2 \left( \frac{\mu^2}{k^2} \right)_{+}
 S(k^2,p^2) = } \nonumber \\
& & \int_{0}^{p^2} dk^2 \,\frac{\mu^2}{k^2} \left[ S(k^2,p^2) -
S(0,p^2) \right] + \int_{p^2}^{\infty} dk^2 \,\frac{\mu^2}{k^2}
S(k^2,p^2)\; .
\ea
\noindent Simply taking
\[G_{in}(k^2) = A \left(\frac{\mu^2}{k^2}\right)_+ \]
as an input function once again leads to a mass term and higher terms in
the expansion Eq.~(8) are necessary. Then we do have the chance of
finding self-consistency for a gluon propagator as singular
as $1/k^4$ and hence {\it confining} quarks. With
\be
G_{in}(p^2) = \left\{ \begin{array}{ll}
A \left({\mu^2}/{p^2} \right)_+ + \left({p^2}/{\mu^2} \right)
& \mbox{if $p^2 < \mu^2 $} \\ \\
1 & \mbox{if $p^2 > \mu^2 $}
\end{array}
\right. \qquad ,\ee

\noindent we find, after mass renormalization:
\be
\frac{1}{G_{out}(p^2)} = 1 + \frac{g^{2}C_{A}}{48\pi^2} \left[
-\frac{479}{24}\frac{p^2}{\mu^2} +
\frac{13}{8}\frac{p^2}{\mu^2}\ln\left(\frac{\mu^2}{p^2}\right)
-\frac{81}{4} -
\frac{25}{4}\ln\left(\frac{\Lambda^2}{\mu^2}\right) \right] \qquad .\ee

\noindent The ultraviolet divergent constant can be arranged to cancel
the 1 and, again,  we find self-consistency. This is the result found
numerically by Brown and Pennington \cite{BP} with a positive infrared
enhancement to the gluon renormalization function, i.e. $A > 0$.
\vspace{0.4cm}

Now we check, whether it is possible in the Landau gauge, to find the
behaviour Cudell and Ross discovered using Schoenmaker's approximation
in the axial gauge. With
\be
G_{in}(p^2) = \left\{ \begin{array}{ll}
\left({p^2}/{\mu^2}\right)^{1-c} & \mbox{if $p^2< \mu^2$} \\ \\
1 & \mbox{if $p^2> \mu^2$}
\end{array}
\right. \ee

\noindent we find, again after mass renormalization:
\ba
\frac{1}{G_{out}(p^2)} &=& 1 + \frac{g^2C_{A}}{48\pi^2} \left[
\ D_{1} + \ D_{2} \left( \frac{\mu^2}{p^2}\right)^{1-c}
+ \ D_{3} \left(\frac{p^2}{\mu^2} \right)^{1-c} +\  D_{4}\left(
\frac{p^2}{\mu^2}\right)^{c} + ...\right]  \ ,
\ea
where
\ba
D_{1} &=& -\left( \frac{3}{2} + \frac{5+6c}{1-c} +\frac{25}{4}\ln\left(
\frac{\Lambda^2}{\mu^2} \right) \right)\nonumber \qquad , \\
D_{2} &=& -\left( \frac{3}{4(2-2c)} +\frac{3}{4}\ln\left(
\frac{\Lambda^2}{\mu^2} \right) \right)\nonumber \qquad , \\
D_{3} &=& -\frac{1971}{60} +\frac{29c}{2} +\frac{37}{20c} +
\frac{6-13c}{2(1-c)} +\frac{59-32c}{4(2-c)}
+\frac{155-64c}{8(3-c)} \nonumber \\
& &+\frac{127-49c}{8(4-c)} +\frac{23-11c}{4(5-c)} -\frac{125+61c}{8(1-2c)}
-\frac{55+6c}{8(2-2c)} +\frac{3}{4(3-2c)} \nonumber \\
& &-8(2-c)\Psi(-2c) -8(2-c)\Psi(1) \phantom{\frac{2}{2}}\nonumber
\qquad , \\
D_{4} &=& \frac{61+6c}{8(1-2c)}\nonumber \qquad .
\ea

\noindent Thus the dominant infrared behaviour is:
\[\frac{1}{G_{out}(p^2)} \rightarrow - \left(\frac{\mu^2}{p^2}
\right)^{1-c}\]
and self-consistency is spoiled by a negative sign, since $c$ is small and
positive.

\vspace{0.8cm}
{\bf {\large 5. Confined Gluons}}
\vspace{0.4cm}

A gluon propagator, which is less singular than $1/k^2$ for $k^2
\rightarrow 0$, and hence describes {\it confined} gluons appears to be a
self-consistent solution only of the axial gauge Schwinger-Dyson
equation using Schoenmaker's approximate integral Eq.~(7). In the
Landau gauge this behaviour of the gluon propagator is not possible~:
a minus-sign spoils self-consistency.
We should therefore comment on the origin of this crucial minus sign.

Starting from BBZ's integral
Eq.~(6), there is no difference in sign between the two gauges.
Eq.~(6) is Bose-symmetric (as it should be) and can therefore
be rewritten as~:
\ba
\lefteqn{\frac{p^2}{F(p^2)}\left(1-\gamma \right) = p^2
\left(1-\gamma \right)} \nonumber \\
& & - \frac{g^2 C_{A}}{2}\,\int\frac{d^4k}{(2\pi)^4}\frac{n\cdot(k-q)}
{n^2}\, \Delta^{\lambda\rho}_{(0)}(k)\Delta^{\lambda\sigma}_{(0)}(q)\,
2k\cdot n \ K_{\rho\sigma}(k,p)
\ea
where
\ba
K_{\rho\sigma}(k,p) =& & \delta_{\rho\sigma}F(q^{2}) -
\frac{F(q^{2})-F(k^{2})}{k^{2}-q^{2}}( \delta_{\rho\sigma} k\cdot q
- k_{\rho}q_{\sigma}) \nonumber  \\
&+&  \frac{ F(k^{2})-F(p^{2})}{p^{2}-k^{2}}\frac{F(q^{2})}
{F(p^{2})}\ (p+k)_{\rho} p_{\sigma} \nonumber
\ea

\noindent whereas, taking the starting equation of Schoenmaker's paper
(Eq~3.5 of Ref. \cite{Schoen})
we find:
\ba
\lefteqn{\frac{p^2}{F(p^2)}\left(1-\gamma \right) = p^2\left(1-\gamma
\right)} \nonumber \\
& & -\frac{i}{2}g^2C_{A}\int\frac{d^4k}{(2\pi)^4}\,\frac{n\cdot(k-q)}
{n^2}\, \Sigma_{\rho\sigma}(k,q)[-2k\cdot n \  K_{\rho\sigma}(k,p)]
\ea
where
\[\Sigma_{\rho\sigma}(k,q) = (i)^2\Delta^{\lambda\rho}_{(0)}(k)
\Delta^{\lambda\sigma}_{(0)}(q) \qquad .\]
Schoenmaker formulates his equation in Minkowski space. Performing a
Wick rotation to transform to Euclidean space by~:
$d^4 k_{M} \rightarrow i d^4 k_{E}$, we
find that Schoenmaker's equation, Eq.~(26), becomes~:
\ba
\lefteqn{\frac{p^2}{F(p^2)}\left(1-\gamma \right) = p^2
\left(1-\gamma \right)} \nonumber \\
& & + \frac{g^2 C_{A}}{2}\,\int\frac{d^4k}{(2\pi)^4}\frac{n\cdot(k-q)}
{n^2}\, \Delta^{\lambda\sigma}_{(0)}(k)\Delta^{\lambda\rho}_{(0)}(q)\,
2k\cdot n \ K_{\sigma\rho}(k,p)
\ea

\noindent which differs from Eq.~(25) by a crucial minus sign.

\vspace{0.3cm}
We therefore see that in the axial gauge using BBZ's
integral equation for the gluon propagator, and
simplifying the angular dependence in the way  Schoenmaker does in order
to make an analytical discussion of the infrared behaviour of the
propagator possible, yields an integral equation very similar to the
one found by Brown and Pennington~\cite{BP} in the Landau gauge. These
equations lead to the correct perturbative behaviour at large momenta.
In contrast a self-consistent solution of the gluon propagator
less singular than $1/k^2$ for $k^2 \rightarrow 0$ cannot be
found in either gauge. Schoenmaker's own equation, which is the
starting point for the study of Cudell and Ross~\cite{CR} for instance,
has an incorrect additional minus sign. This should have been heralded by the
self-consistent enhanced gluon of Eq.~(11) having a negative sign, using
Schoenmaker's equation. In an axial gauge this sign should have been a little
worrying for a wavefunction renormalization of a state with positive definite
norm.

\vspace{0.8cm}
{\bf {\large 6. Summary and Conclusion}}
\vspace{0.4cm}

We have studied the Schwinger-Dyson equation of the gluon propagator to
determine analytically the possible infrared solutions for the gluon
renormalization function $G(p^2)$.
In both the axial and Landau gauges, one can find a self-consistent
solution, which behaves as $1/k^2$ for $k^2 \rightarrow 0$ and hence a
propagator which is as singular as $1/k^4$ for $k^2 \rightarrow 0$.
This form of the gluon propagator is consistent with area law behaviour
of the Wilson loop, which is regarded as a signal for confinement
\footnotemark.
\footnotetext{Though as remarked at the end of Sect.~4, axial gauge
studies are seriously marred by the simplifying assumption that
$H \equiv 0$ in Eq.~(1).}
Numerical studies \cite{BBZ}, \cite{BP} have shown that a gluon
propagator with such an enhanced behaviour in the infrared region that
connects to the perturbative regime at a finite momentum (as indicated
by experiment) can indeed be found as a self-consistent solution to the
boson Schwinger-Dyson equation. Such a behaviour of the boson propagator has
been shown to give quark propagators with no physical poles \cite{Stuller}.
Furthermore, extending these
non-perturbative methods to hadron physics, it has been found that
a regularized, infrared singular gluon propagator together with the
Schwinger-Dyson equation for the quark self-energy, gives rise to a
good description of dynamical chiral symmetry breaking.  For instance,
one obtains values for quantities such as the pion decay constant that
agree with experimental results~\cite{SDE}.

\vspace{0.3cm}
A gluon propagator which is less singular than $1/k^2$ for $k^2
\rightarrow 0$, and hence describes {\it confined} gluons, cannot be found in
either the axial or the Landau gauge. Solutions of this type have only
been found using approximations to the gluon Schwinger-Dyson equation
with an incorrect sign. Possible consequences for models of the pomeron
are discussed elsewhere \cite{pomeron}.

We should also mention the related work of the group of Stingl et
al.~\cite{Stingl}.  They too start from an approximate, but larger,
set of Schwinger-Dyson equations, which is then to be solved
self-consistently. However, the method employed is completely different.
The philosophy~\cite{Stingl} is to obtain the solution of these
equations as  power series in the coupling, as in perturbation theory,
and to include non-perturbative effects by letting each Green's function
depend upon a spontaneously generated mass scale, $b(g^2)$.
The gluon propagator is assumed to be of the form (see Fig.~1)

\be \Delta(k^2) \equiv \frac{G(k^2)}{k^2} = \frac{k^2}{k^4+b^4} \qquad ,\ee

\noindent representing {\it confined} gluons.  This grossly violates the
masslessness condition of Eq.~(8).  In general, gluon masses can only
arise in 4-dimensions if the vertex functions have singularities
themselves corresponding to coloured massless scalar states, otherwise
the Slavnov-Taylor identities sufficiently constrain the vertex functions
to require the inverse of the gluon propagator to vanish at $k
\rightarrow 0$ \footnotemark.
\footnotetext{This remark equally applies to the work of Zwanziger
\cite{Zwanziger} and lattice studies by \cite{lattice}.}
Not only do the vertices of Stingl et al.  have these massless singularities
but self-consistency can only be found if the 3-gluon vertex is complex,
when conventional understanding of its singularity structure would lead us to
expect it to be real for momenta which in Minkowski space are spacelike.
Subsequently Hawes, Roberts and Williams \cite{HRW} and Alkofer and Bender
\cite{AB}
have shown that with this gluon propagator, Eq.~(28), the solution of the
quark
Schwinger-Dyson equation does not describe a confined particle. They
therefore also
conclude that the full gluon propagator in QCD cannot vanish in the infrared
region.

\vspace{1cm}
\noindent To summarise:

At first sight there appears to be a distinction between a {\it confining} and
a {\it confined} gluon. A {\it confining} gluon is one whose interactions lead
to quark confinement. $\Delta(k^2) \sim 1/k^4$ behaviour is of this
{\it confining} type. In contrast, it is sometimes argued that $\Delta(k^2)$
must be less singular than $1/k^2$ to ensure that gluons themselves do
not propagate over large distances. However, whether gluons are {\it confining}
or {\it confined} are not real alternatives. Gluons must be both. They confine
quarks by having very strong long range interactions. They themselves
are confined by not having a Lehman representation that any physical
asymptotic state must have \cite{SDE}.

While infrared singular gluons satisfy both criteria, softened gluons
though {\it confined}, do not generate quark confinement or dynamical chiral
symmetry breaking, which are features of our world. Remarkably, a study
of the field equations of QCD  reveals this theory naturally exhibits
these aspects with an infrared enhanced gluon propagator.

\vspace{0.8cm}
{\bf {\large Acknowledgements}}
\vspace{0.4cm}

KB thanks the University of Durham for the award of a research
studentship. We are grateful to Douglas Ross for repeating our
calculations and confirming that the sign in Schoenmaker's Eq.~(3.5) is
incorrect. Many of the underlying ideas discussed here were formulated by
MRP in collaboration with Nick Brown --- his memory lingers on.

\end{document}